\documentclass[ALICE,manyauthors]{cernphprep}
\usepackage[comma,square,numbers,sort&compress]{natbib}
\usepackage{hyperref} 
\usepackage{lineno}
\usepackage{color}
\usepackage{caption}

\usepackage{graphicx}
\usepackage{epstopdf}
\usepackage{amssymb}
\usepackage{xspace}
\usepackage{amsmath}
\usepackage{upgreek}
\usepackage{hyperref}
\usepackage{doi}
\usepackage{multirow}
\usepackage{makecell}
\usepackage{booktabs}

\usepackage[T1]{fontenc}
\usepackage{orcidlink}

\DeclareRobustCommand{\unit}[2][]{%
        \begingroup%
                \def\0{#1}%
                \expandafter%
        \endgroup%
        \ifx\0\@empty%
                \ensuremath{\mathrm{#2}}%
        \else%
                \ensuremath{#1\,\mathrm{#2}}%
        \fi%
        }
\DeclareRobustCommand{\unitfrac}[3][]{%
        \begingroup%
                \def\0{#1}%
                \expandafter%
        \endgroup%
        \ifx\0\@empty%
                \raisebox{0.98ex}{\ensuremath{\mathrm{\scriptstyle#2}}}%
                \nobreak\hspace{-0.15em}\ensuremath{/}\nobreak\hspace{-0.12em}%
                \raisebox{-0.58ex}{\ensuremath{\mathrm{\scriptstyle#3}}}%
        \else
                \ensuremath{#1}\,%
                \raisebox{0.98ex}{\ensuremath{\mathrm{\scriptstyle#2}}}%
                \nobreak\hspace{-0.15em}\ensuremath{/}\nobreak\hspace{-0.12em}%
                \raisebox{-0.58ex}{\ensuremath{\mathrm{\scriptstyle#3}}}%
        \fi%
}

\newlength{\smallerpicsize}
\newlength{\smallpicsize}
\newlength{\mediumpicsize}
\newlength{\largepicsize}
\topfigrule
\newcommand {\meanpT}    {\ensuremath{\langle p_{\mathrm{T}} \kern-0.1em\rangle}\xspace}
\newcommand {\mean}[1]   {\ensuremath{\langle #1 \kern-0.1em\rangle}\xspace}

\newcommand {\ee}        {\ensuremath{\mathrm{e^+e^-}}\xspace}
\newcommand {\mumu}      {\ensuremath{\upmu^+\upmu^-}\xspace}

\newcommand{\dEdx}       {\ensuremath{\mathrm{d}E/\mathrm{d}x}\xspace}

\newcommand {\ep}        {\mbox{$\mathrm {e\kern-0.05em p}$}\xspace}
\newcommand {\pp}        {\mbox{$\mathrm {p\kern-0.05em p}$}\xspace}

\newcommand {\ppBoldMath} {\mbox{$\mathrm { \mathbf p\kern-0.05em \mathbf p }$}\xspace}

\newcommand {\MeanNpart} {\mbox{\ensuremath{< \kern-0.15em N_{part} \kern-0.15em >}}}

\newcommand {\mass}         {\mbox\mathrm{MeV$\kern-0.15em /\kern-0.12em c^2$}}
\newcommand {\tev}          {\ensuremath{\mathrm{TeV}}\xspace}
\newcommand {\gev}          {\ensuremath{\mathrm{GeV}}\xspace}
\newcommand {\mev}          {\ensuremath{\mathrm{MeV}}\xspace}

\newcommand {\mmom}         {\ensuremath{\mathrm{MeV\kern-0.15em /\kern-0.12em c}}\xspace}
\newcommand {\gmom}         {\ensuremath{\mathrm{GeV\kern-0.15em /\kern-0.12em c}}\xspace}
\newcommand {\mmass}        {\ensuremath{\mathrm{MeV\kern-0.15em /\kern-0.12em c^2}}\xspace}
\newcommand {\gmass}        {\ensuremath{\mathrm{GeV\kern-0.15em /\kern-0.12em c^2}}\xspace}

\newcommand {\fm}           {\ensuremath{\mathrm{fm}}\xspace}

\newcommand {\dg}           {\ensuremath{\kern+0.1em ^\circ}\xspace}

\newcommand{\pt}            {\ensuremath{p_\mathrm{T}}\xspace}

\newcommand{\Ph}           {\ensuremath{\upphi}\xspace}

\newcommand{\Kstar}         {\ensuremath{\mathrm{K}^{*0}}\xspace}

\newcommand{\Jpsi}          {\ensuremath{\mathrm{J}/\uppsi}\xspace}

\newcommand{\DstartoDpi}    {\mbox{\ensuremath{\mathrm{D}^{*+} \to \mathrm{D}^0 (\to \mathrm{K}^-\uppi^+) \uppi^+}}\xspace}

\newcommand{\Dzero}         {\ensuremath{\mathrm{D^0}}\xspace}

\newcommand{\Dstar}         {\ensuremath{\mathrm{D^{*+}}}\xspace}

\newcommand{\Lc}            {\ensuremath{\mathrm{\Lambda_{c}^+}}\xspace}

\newcommand{\ccbar}         {\ensuremath{\mathrm{c\overline{c}}}\xspace}
\newcommand{\bbbar}         {\ensuremath{\mathrm{b\overline{b}}}\xspace}

\newcommand{\GeV}           {\gev}
\newcommand{\MeV}           {\mev}

\newcommand{\TeV}           {\tev}

\newcommand{\mum}           {\ensuremath{\mathrm{\upmu m}}\xspace}

\newcommand{\fprompt}       {\ensuremath{f_\mathrm{prompt}}\xspace}

\newcommand{\AccEff}        {\ensuremath{(\mathrm{Acc}\times\epsilon)}\xspace}
\newcommand{\effP}[1]       {\ensuremath{(\mathrm{Acc}\times\epsilon)_{\mathrm{prompt}, #1}}\xspace}
\newcommand{\effNP}[1]      {\ensuremath{(\mathrm{Acc}\times\epsilon)_{\mathrm{non\text{-}prompt},#1}}\xspace}
\newcommand{\rawY}[1]       {\ensuremath{Y_#1}\xspace}

\newcommand{\rzz}           {\ensuremath{\rho_{00}}\xspace}
\newcommand{\rzzp}          {\ensuremath{\rho_{00}^\mathrm{prompt}}\xspace}
\newcommand{\rzznp}         {\ensuremath{\rho_{00}^\mathrm{non\text{-}prompt}}\xspace}
\newcommand{\rzzobs}         {\ensuremath{\rho_{00}^\mathrm{obs}}\xspace}
\newcommand{\cost}          {\ensuremath{\cos{\vartheta^*}}\xspace}
\newcommand{\lt}            {\ensuremath{\lambda_\vartheta}\xspace}
\newcommand{\lp}            {\ensuremath{\lambda_\varphi}\xspace}
\newcommand{\ltp}           {\ensuremath{\lambda_{\vartheta\varphi}}\xspace}
\newcommand{\Np}               {\ensuremath{N_\mathrm{prompt}}\xspace}
\newcommand{\Nnp}              {\ensuremath{N_\mathrm{non\text{-}prompt}}\xspace}

\newcommand{\snn}             {\ensuremath{\sqrt{s_{\mathrm{NN}}}}\xspace}

\newcommand{\dstar}           {\ensuremath{\mathrm{D^{*+}}}\xspace}

\newcommand{\pythia}          {\textsc{Pythia~8}\xspace}

\newcommand{\pythiaprec}      {\textsc{Pythia~8.243}\xspace}

\newcommand{\evtgen}          {\textsc{EvtGen}\xspace}
\newcommand{\geant}           {\textsc{Geant~3}\xspace}
\newcommand{\hijing}          {\textsc{Hijing~1.36}\xspace}

\begin{document}%
\begin{titlepage}

\PHyear{2025}
\PHnumber{072}      % required, will be obtained from PH
\PHdate{26 March}  % required, will be obtained from PH %
  
  %%% Put your own title + short title here:
\title{First measurement of $\pmb{\dstar}$ vector meson spin alignment in Pb--Pb collisions at $\pmb{\snn = 5.02}$ TeV}   
\ShortTitle{Spin alignment of \dstar vector mesons in Pb--Pb collisions at $\snn = 5.02$ TeV}
% appears on right page headers  

%%% Do not change the next lines
\Collaboration{ALICE Collaboration\thanks{See Appendix~\ref{app:collab} for the list of Collaboration members}}
\ShortAuthor{ALICE Collaboration} 

\begin{abstract}
  The first measurement of prompt \Dstar-meson spin alignment in ultrarelativistic heavy-ion collisions with respect to the direction orthogonal to the reaction plane is presented. 
  The spin alignment is quantified by measuring the element \rzz of the diagonal spin-density matrix for prompt \Dstar mesons with $4<\pt<30~\GeV/c$ in two rapidity intervals, $|y|<0.3$ and $0.3<|y|<0.8$, in central (0--10\%) and midcentral (30--50\%) Pb--Pb collisions at $\snn=5.02~\TeV$. 
  Evidence of spin alignment $\rzz>1/3$ has been found for $\pt>15~\GeV/c$ and $0.3<|y|<0.8$ with a significance of $3.1\sigma$. 
  The measured spin alignment of prompt \Dstar mesons is compared with the one of inclusive \Jpsi mesons measured at forward rapidity ($2.5 < y < 4$). 
\end{abstract}
\end{titlepage}
\setcounter{page}{2}

\section{Introduction}

High-energy heavy-ion collision experiments study the strongly interacting quantum chromodynamics (QCD) matter at very high temperature and extreme energy densities, where a phase of deconfined quarks and gluons called quark--gluon plasma (QGP) is expected to be formed~\cite{ALICE:2022wpn}. 
Theoretical studies, alongside experimental evidences, have shown that the QGP behaves as a strongly interacting fluid. Several properties of the QGP medium, such as shear viscosity to entropy-density ratio, stopping power, and diffusion coefficients, have been measured, leading to the discovery of an emergent property of QCD matter that characterises it as a nearly perfect fluid~\cite{ALICE:2022wpn}.
Recently there has been a thoughtful understanding of the QGP properties in systems possessing large angular momentum and/or immersed in a strong magnetic field. 
In non-central heavy-ion collisions with a non-zero impact parameter, defined as the distance between the centers of the two colliding nuclei, a large angular momentum of $O (10^{7}~\hslash)$~\cite{THAM} and a strong magnetic field of $O(10^{15}~\mathrm{T})$~\cite{THMF} are expected to be created at LHC energies in the direction perpendicular to the reaction plane, defined by the direction of the impact parameter of the two nuclei and the beam direction. 
During the collision, a sizable fraction of the total angular momentum is transferred to the created QGP fluid and the remaining fraction is carried away by the spectator nucleons that do not participate in the collision. The angular momentum can be transferred to the QGP in the form of fluid vorticity along the direction of the angular momentum.
In general, the relation between angular momentum and fluid vorticity is not straightforward and depends on the local fluid density that affects the moment of inertia through a complex dependence in case of an expanding system.
The theory calculation reported in Ref.~\cite{Jiang:2016woz} suggests that while the deposited angular momentum in the QGP medium is conserved in time, the fluid vorticity strongly depends on the time evolution of the fluid, in particular the fluid vorticity is significantly higher in the initial stage of the QGP evolution. As opposed to the angular momentum, the magnetic field is short lived (a fraction of fm/$c$) and its lifetime depends on the conductivity of the medium~\cite{Zhang:2023ppo}. In the presence of fluid vorticity and an intense magnetic field, quarks in the QGP would be polarised and their polarisation can be further transferred to the final state hadrons during the process of hadronisation.

The polarisation of a vector meson is challenging to measure in strong hadronic decays~\cite{Yang:2017sdk}. However information about it can be inferred via the measurement of the spin alignment, which is defined by the deviation of the spin density matrix element~\cite{Yang:2017sdk} \rzz from $1/3$. The value of \rzz implies the probability of finding a vector meson in the state with spin projection zero out of possible states with spin projection $-1$, 0, and 1~\cite{Fano:1957zz}. If the spin of the particles is oriented isotropically, then $\rzz = 1/3$. The spin alignment is studied by measuring the angular distribution of the decay products of vector mesons in the rest frame of the decaying vector meson and a quantisation/polarisation axis  which in heavy-ion collisions can be defined as the perpendicular to the reaction plane. The angular distribution of the vector meson decays to two spin zero hadrons is expressed as
 \begin{equation}
 \frac{\mathrm{d}N}{\mathrm{d}\cos{\vartheta^{*}}} \propto [ 1 - \rzz + (3\rzz - 1 )\cos^{2}{\vartheta^{*}} ],
 \label{eq:dndcost}
 \end{equation}
 where $\vartheta^{*}$ is the angle between the momentum of one of the decay daughters in the rest frame of the vector meson with respect to the quantisation axis~\cite{Schilling:1969um}. While in proton--proton (pp) collisions the helicity axis, i.e. the direction of the decaying hadron in the laboratory frame, is used as the quantisation axis, in heavy-ion collisions the direction defined by the angular momentum and the magnetic field vectors is used. The directions of these two quantisation axes are correlated through the anisotropic collective expansion observed in heavy--ion collisions, which is quantified by coefficients in a Fourier decomposition of the azimuthal-angle distribution of final-state particle momenta, with the second-harmonic coefficient $v_2$, called elliptic flow~\cite{Snellings:2011sz}.
 First evidence of light-flavour vector meson (\Kstar and \Ph) spin alignment in heavy-ion collisions was reported by the ALICE Collaboration~\cite{ALICE:2019aid} and in later years \Ph meson spin alignment was also observed by the STAR Collaboration~\cite{STAR:2022fan}. At LHC energies, a negative deviation ($< 1/3$) of the parameter \rzz for \Kstar and \Ph mesons was measured, whereas at RHIC energies a positive deviation ($> 1/3$) of \rzz for \Ph mesons was observed. The measured spin alignment of light-flavour vector mesons at RHIC and LHC energies are surprisingly much larger compared to expectations from the measurement of global $\Lambda$ polarisation in the non-relativistic thermodynamic limit considering the vorticity as the only source of polarisation~\cite{ALICE:2019aid, Becattini:2016gvu}. This triggered new theory developments to understand the origin of vector meson spin alignment. Although the hyperon polarisation is sensitive to the average quark polarisation, the spin alignment of vector mesons is sensitive to the local fluctuation or correlation of quark--antiquark spin~\cite{Lv:2024uev}. On one hand, theoretical studies based on local polarisation originating from the anisotropic expansion of the fireball~\cite{local1, local2}, fluctuations of quark polarisation in turbulent colour fields due to locally fluctuating axial charge currents~\cite{Muller:2021hpe}, and local spin correlation induced by the colour fields in the glasma phase~\cite{Kumar:2023ghs} showed that these mechanisms could lead to large vector meson spin alignment. In Refs.~\cite{Liang:2004xn, Yang:2017sdk} hadronisation via coalescence (recombination of polarised quarks) is predicted to give $\rzz < 1/3$, however its deviation from $1/3$ is expected to be very small compared to the reported measurements. On the other hand, the observed $\rzz > 1/3$ for \Ph mesons at RHIC energies can be qualitatively explained by models that incorporate the fluctuation of a \Ph meson field~\cite{Sheng:2019kmk, Sheng:2022wsy, Sheng:2023urn}. In this model, the strange and antistrange quarks can be polarised by the \Ph meson field induced by the current of pseudoscalar bosons when they form the \Ph meson. The local correlation or fluctuation of the vector-meson field leads to $\rzz > 1/3$ for \Ph mesons produced from the coalescence of polarised strange and antistrange quarks. Furthermore, recent calculations based on the holographic approach also predict a significant spin alignment of flavourless vector mesons due to the modification of the spectral function for different spin channels in the presence of thermal background~\cite{Sheng:2024kgg}. Calculations based on the holographic approach are also qualitatively consistent with the \Ph meson spin alignment measurements at RHIC energies. Despite the several theoretical approaches, a single consistent picture to describe the light flavour vector meson spin alignment across different collision energies and particles does not exist yet and requires further theoretical developments.

Unlike light-flavour hadrons, which originate at a late stage of a heavy-ion collision, the heavy-flavour hadrons could be more sensitive to the initial stages as the charm quarks are predominantly produced in the initial hard scattering processes with typical production time of $O(0.1~\fm/c)$, which is lower than the QGP formation time $O(1~\fm/c)$~\cite{ALICE:2022wpn}. Therefore, charm quarks are expected to be affected more by the magnetic field and vorticity, which are significantly larger at the time of charm-quark formation~\cite{Das:2016cwd}. Additionally, high transverse-momentum (\pt) charm quarks are likely to retain a greater degree of this polarisation due to their shorter interaction time with the medium constituents, also depending on their spin-relaxation time~\cite{Dey:2025ail}. The ALICE Collaboration has recently measured a significant spin alignment of inclusive \Jpsi vector mesons~\cite{ALICE:2022dyy} at low \pt and at forward rapidity which is qualitatively consistent with the expectation from the calculation based on the holographic approach reported in Ref.~\cite{Sheng:2024kgg}. The \rzz parameter of \Dstar mesons has been measured in pp collisions by the ALICE Collaboration for both prompt (directly originating from charm-quark hadronisation, or the decay of excited charm hadron decays) and non-prompt (from beauty-hadron decays) production~\cite{ALICE:2022byg}. In the case of prompt production, no significant spin alignment was observed, while in the case of non-prompt \Dstar mesons a $\rzz > 1/3$ was found, as a consequence of the helicity conservation in weak decays~\cite{ALICE:2022byg}. 

In this paper, we report the first measurement of prompt open charm hadron (\Dstar) spin alignment at midrapidity ($|y|$ $<$ 0.8) in heavy--ion collisions. The measurement of the \rzz of prompt \Dstar mesons in Pb--Pb collisions represents useful input for the modeling of charm-hadron spin alignment in heavy-ion collisions and can shed light on the propagation of charm quarks under extreme magnetic fields and vorticity expected at the initial stages of heavy-ion collisions. In addition, measurements at high \pt are important to understand how the quark polarisation is propagated to final-state hadrons when the underlying hadronisation process is quark fragmentation. The organisation of the article is as follows. The experimental apparatus, data analysis technique, sources of systematic uncertainties are described in Secs.~\ref{sec:analysis}, ~\ref{sec:data}, ~\ref{sec:syst}, respectively. Results are shown in Sec.~\ref{sec:res}, and the article is finally summarised in Sec.~\ref{sec:sum}.

\section{Experimental apparatus}
\label{sec:analysis}

The ALICE apparatus comprises a central barrel, which is composed of a set of detectors for charged-particle reconstruction and identification at midrapidity, a forward muon spectrometer, and various forward and backward detectors for triggering and event characterisation.
A detailed description of the detectors and an overview of their typical performances can be found in Refs.~\cite{Aamodt:2008zz,ALICE:2014sbx}.
The main detectors used for the analysis presented in this paper are the Inner Tracking System (ITS), a six-layer silicon detector used to track charged particles and for the reconstruction of primary and secondary vertices; the Time Projection Chamber (TPC), which provides track reconstruction as well as particle identification via the measurement of the specific ionisation energy loss \dEdx; and the Time-Of-Flight (TOF) detector, an array of Multigap Resistive Plate Chambers that provides particle identification via the measurement of the
flight time of the particles. These detectors cover the pseudorapidity interval $|\eta| < 0.9$ and are located in a large solenoidal magnet providing a uniform magnetic field of $0.5$ T parallel to the LHC beam direction. In addition, the V0 detector, which consists of two arrays of 32 scintillators each, covering the full azimuth in the pseudorapidity intervals $-3.7 < \eta < -1.7$ (V0C) and $2.8 < \eta < 5.1$ (V0A), and the Zero Degree Calorimeters (ZDC), located at $112.5$ m from the interaction point on either side, were used for event selection and classification.

The analysis was performed using samples of Pb--Pb collisions recorded in 2018 during the second LHC data-taking period with a minimum bias trigger which required coincident signals in the V0A and V0C detectors. Two additional trigger classes were used to enrich the sample of central and midcentral collisions via an online event selection based on the V0-signal amplitude. In order to have a uniform acceptance in pseudorapidity, only events with a primary vertex reconstructed within $\pm10$~cm from the centre of the detector along the beam-line direction were considered in the analysis. 
Collisions were classified into centrality intervals, defined in terms of percentiles of the hadronic Pb--Pb cross section, based on the V0 signal amplitude, as described in detail in Ref.~\cite{ALICE:2015juo}. Background events due to the interaction of one of the beams with residual gas in the vacuum tube and other machine-induced backgrounds were rejected offline using the V0 and the ZDC timing information~\cite{ALICE:2014sbx}.
The samples of central and midcentral collisions consist of approximately $100 \times 10^6$ and $85 \times 10^6$ events in the 0--10\% and 30--50\% centrality intervals, corresponding to integrated luminosities of $\mathcal{L}_{\rm int}~\simeq 130~\upmu \mathrm{b}^{-1}$ and $\mathcal{L}_{\rm int}~\simeq 56~\upmu \mathrm{b}^{-1}$, respectively.

\section{Data analysis}
\label{sec:data}
The V0 detectors were used to determine the second-order harmonic event plane $\psi_2$, which is chosen as an estimator for the reaction plane of the collision. The $\psi_2$ angle is defined as

\begin{equation}
  \psi_2 = \dfrac{1}{2} \tan^{-1} \left(\dfrac{Q_{2,y}}{Q_{2,x}}\right),
  \label{eq:evplane}
\end{equation}

where $Q_{2,x}$ and $Q_{2,y}$ are the second-harmonic flow-vector components computed as

\begin{equation}
	Q_{2,x}^\text{V0} = \sum_{k=1}^{N_\mathrm{cells}} w_k\cos (2\varphi_k),\quad Q_{2,y}^\text{V0} = \sum_{i=k}^{N_\mathrm{cells}} w_k\sin (2\varphi_k).
	\label{eq:qvecV0}
\end{equation}

In the above formula, $N_\mathrm{cells}$ corresponds to the 64 cells of the full V0 detector, $\varphi_k$ is the azimuthal angle of the centre of the cell $k$ and $w_k$ is the amplitude of the signal in cell $k$, once the gain of the single channels is equalised in each ring and the recentering is applied to correct effects of non-uniform acceptance~\cite{Selyuzhenkov:2007zi}.
The event-plane resolution $R_2$, defined as
\begin{equation}
	R_2 = \sqrt{\frac{\langle\cos\left[2(\psi_2^\mathrm{V0}-\psi_2^\mathrm{TPC~\eta>0})\right]\rangle\langle\cos\left[2(\psi_2^\mathrm{V0}-\psi_2^\mathrm{TPC~\eta<0})\right]\rangle}{\langle\cos\left[2(\psi_2^\mathrm{TPC~\eta>0}-\psi_2^\mathrm{TPC~\eta<0})\right]\rangle}},
	\label{eq:evplresol}
\end{equation}

was determined by correlating three sub-events of charged particles reconstructed in the V0 itself, in the positive ($0 < \eta < 0.8$) and negative ($-0.8 < \eta < 0$) semivolumes of the TPC. In the formula in Eq.~\ref{eq:evplresol}, $\psi_2^\mathrm{V0}$ is the second-order harmonic event plane angle determined with the V0 detector, while $\psi_2^\mathrm{TPC~\eta>0}$ and $\psi_2^\mathrm{TPC~\eta<0}$ are those computed using charged-particle tracks reconstructed in the TPC. The brackets $\langle\rangle$ denote the average over all the events within a centrality class. The event-plane resolution obtained with the above formula was about $R_2 = 0.62$ for the 0--10\% centrality class and $R_2 = 0.77$ for the 30--50\% centrality class.

The \Dstar mesons and their charge conjugates were measured at midrapidity ($|y|<0.8$) via the \\\DstartoDpi decay channel, with branching ratio $\mathrm{BR} = (2.67\pm0.03)\%$~\cite{ParticleDataGroup:2024cfk}. The \Dzero-decay candidates were defined combining pairs of tracks reconstructed with the ITS and TPC detectors having the expected charge combinations. Each track was required to have $|\eta| < 0.8$, $\pt > 0.3~\gev/c$, at least 70 (out of 159) associated space points in the TPC, and a minimum of two hits in the ITS, with at least one in either of the two innermost layers to ensure a good pointing resolution. An additional low-\pt track having $|\eta| < 0.8$, $\pt > 50~\mev/c$, and at least two hits in the ITS was then added to form \Dstar-meson candidates.
The analysis was based on the reconstruction of decay-vertex topologies of \Dzero mesons displaced from the interaction vertex. In particular, the proper decay length of \Dzero mesons of $c\tau \approx 123~\mum$ and that of beauty hadrons of $c\tau \approx 500~\mum$ were exploited to resolve the \Dzero-meson decay vertices. In order to reduce the large combinatorial background and to separate the contribution of \Dstar mesons originating from charm hadronisation or excited charm-hadron decays (prompt) from those stemming from beauty-hadron decays (non-prompt), a multiclass classification algorithm based on Boosted Decision Trees (BDTs) was used~\cite{xgboost,hipe4ml}. The variables used to train the BDT algorithm to distinguish among prompt, non-prompt \Dstar mesons and background candidates were based on i) the distance between the reconstructed \Dzero-meson decay vertex and the primary vertex, ii) the \Dzero-meson distance of closest approach to the interaction vertex, iii) the cosine of the pointing angle between the \Dzero-meson candidate line of flight and its reconstructed momentum vector, and iv) the particle identification (PID) information of the decay tracks. The PID information was provided by the specific energy loss and the flight time of particles measured with the TPC and TOF detectors, respectively. Given the three classes of candidates, the BDT output scores are related to the probability of each candidate to be a prompt \Dstar meson, a non-prompt \Dstar meson, or combinatorial background.

Signal samples of prompt and non-prompt \Dstar mesons for the BDT training were obtained from Monte Carlo simulations based on the \hijing~\cite{Wang:1991hta} event generator. In each simulated event, additional \ccbar- and \bbbar-quark pairs were injected using the \pythiaprec event generator~\cite{Sjostrand:2014zea} (Monash-13 tune~\cite{Skands:2014pea}) and \Dstar mesons were forced to decay into the hadronic channel of interest for the analysis. The generated particles were propagated through the detector using the \geant transport package~\cite{GEANT}. The conditions of all the ALICE detectors in terms of active channels, gain, noise level, and alignment, and their evolution with time during the data taking period, were taken into account in the simulations. The background samples were obtained from data in the sideband region of the invariant-mass distribution $\Delta M = M(\mathrm{K}\uppi\uppi) - M(\mathrm{K}\uppi)$. The sideband region was chosen as the invariant-mass interval $ 150 < \Delta M < 170~\MeV/c^{2}$, where no \Dstar signal is present. Independent BDT models were trained for the different \pt intervals and centrality classes of the analysis.

\begin{figure}[!t]
\begin{center}
\includegraphics[width=1.\textwidth]{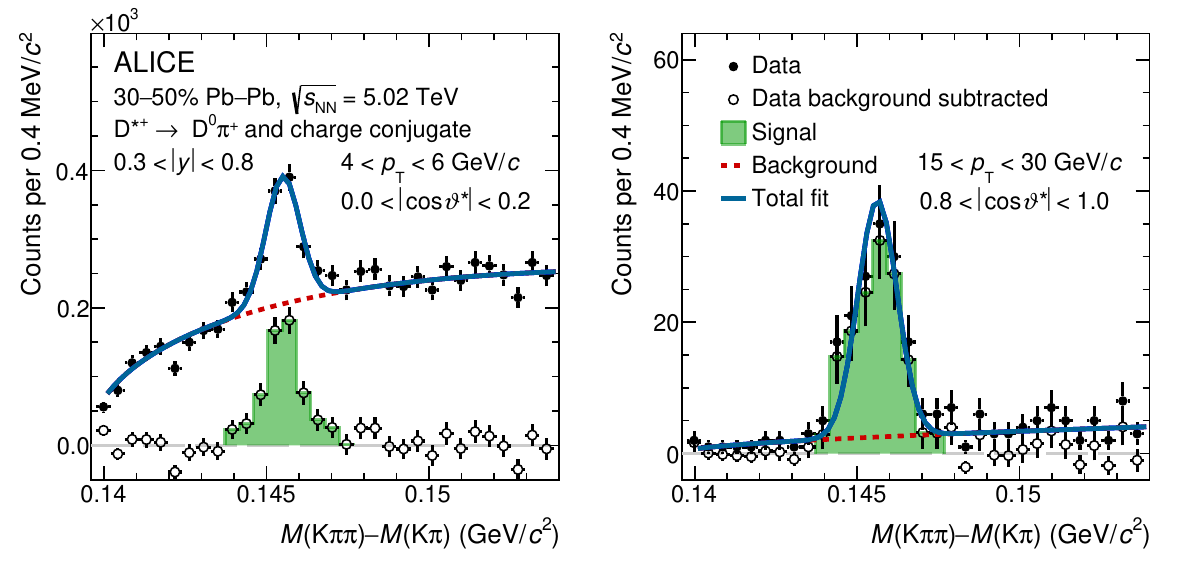}
\caption{Invariant-mass distributions $\Delta M$ of \Dstar candidates with $4<\pt<6~\GeV/c$ and $0.0<|\cost|<0.2$ (left panel), and with $15<\pt<30~\GeV/c$ and $0.8<|\cost|<1.0$ (right panel) in the rapidity interval $0.3<|y|<0.8$ for the 30--50\% centrality class.}
\label{fig:invmass}
\end{center}
\end{figure}

The analysis was performed in five \pt intervals within the $4<\pt<30~\GeV/c$ range, and in the 0--10\% and 30--50\% centrality classes. Furthermore, the \Dstar candidates were divided into two rapidity intervals, $|y|<0.3$ and $0.3<|y|<0.8$. A selection on the BDT score related to the probability to be background was applied in order to suppress the large combinatorial background, while a selection on the BDT score related to the probability to be a prompt \Dstar meson was required to reduce the contamination from non-prompt \Dstar mesons. For each centrality, \pt, and rapidity range, the raw yield of \Dstar mesons was extracted with a fit to the distribution of the invariant mass $\Delta M$ in five $|\cost|$ intervals, $\vartheta^*$ being the angle between the momentum of either the \Dzero-meson or the pion daughters in the rest frame of the \Dstar meson with respect to the quantisation axis, defined as the direction orthogonal to the plane determined by the $\pmb{Q}_2$ flow vector and the beam axis. The $\Delta M$ distribution was fitted with a combination of the Gaussian function corresponding to the \Dstar signal and a background function. The shape of the background distribution was described with the function $p_0\sqrt{\Delta M - M_{\uppi}}\text{ }e^{p_1(\Delta M - M_{\uppi})}$, where $p_0$ and $p_1$ are free parameters and $M_{\uppi}$ is the charged-pion rest mass. The signal was evaluated with a bin counting method, i.e. as the integral of the counts in the invariant-mass distribution within $\pm 3\sigma$, where $\sigma$ is the Gaussian width, after subtracting the fitted background function. Figure~\ref{fig:invmass} shows two examples of fits to the $\Delta M$ distributions of \Dstar candidates with $4<\pt<6~\GeV/c$ and $0.0<|\cost|<0.2$ (left panel), and with $15<\pt<30~\GeV/c$ and $0.8<|\cost|<1.0$ (right panel) in the rapidity interval $0.3<|y|<0.8$ for the 30--50\% centrality class. The width of the Gaussian function for the \Dstar signal was fixed to the value from the MC simulation to improve the stability of the fits.

\begin{figure}[!t]
\begin{center}
\includegraphics[width=1.\textwidth]{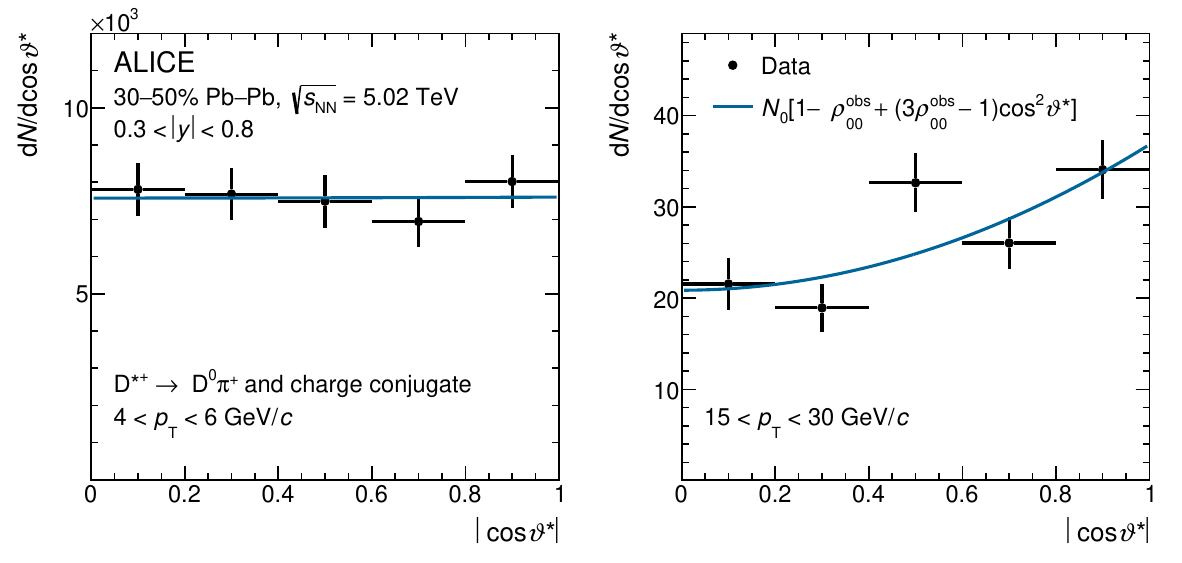}
\caption{Acceptance and efficiency corrected angular distributions of the decaying pion in the rest frame of the \Dstar candidates with $4<\pt<6~\GeV/c$ and with $15<\pt<30~\GeV/c$ in the rapidity interval $0.3<|y|<0.8$ for Pb--Pb collisions in the 30--50\% centrality class. The distributions are fitted with the function from Eq.~\ref{eq:dndcost}. Only statistical uncertainties are shown.}
\label{fig:costdist}
\end{center}
\end{figure}

The raw yields were corrected for the product of the geometrical acceptance and the reconstruction and selection (including BDT) efficiency factors \AccEff for \Dstar mesons in each \cost and \pt intervals. The MC simulations used to compute these correction factors are analogous of those adopted for the BDT trainings. The impact on the \rzz measurement due to the shape of the \cost and \pt distributions in the MC simulations was studied by weighting the generated MC distributions in order to reproduce the measured \cost and \pt distributions. The variation of the estimated \AccEff values was found to be smaller than 0.1\% in the analysed \pt intervals. The corrected angular distributions of \Dstar mesons with $4<\pt<6~\GeV/c$ and $15 < \pt < 30~\GeV/c$ in the $0.3<|y|<0.8$ rapidity interval for the 30--50\% Pb--Pb collisions are shown in the left and right panels of Fig.~\ref{fig:costdist}, respectively. The absolute value of \cost was used considering the symmetry with respect to zero expected for the angular distribution (see Eq.~\ref{eq:dndcost}) and  the limited size of the analysed data sample. The angular distributions were fitted with the functional form given in Eq.~\ref{eq:dndcost} to extract the \rzzobs values for each \pt, rapidity, and centrality interval.

In order to obtain the \rzz parameter with respect to the direction orthogonal to the reaction plane, a correction due to the finite resolution of the second-harmonic event plane $R_2$ has to be applied~\cite{Tang:2018qtu,ALICE:2019aid}. In particular, the corrected \rzz parameter can be computed as

\begin{equation}
\rzz = \frac{1}{3} + \left(\rzzobs - \frac{1}{3}\right) \cdot \frac{4}{1 + 3 R_2},
\end{equation}

where \rzzobs is the value extracted from the acceptance and efficiency corrected \cost distribution of \Dstar mesons.

The measured raw yields contain a mixture of prompt and non-prompt \Dstar mesons. The extracted \rzz parameters are therefore a linear combination of prompt (\rzzp) and non-prompt (\rzznp) contributions, and can be expressed as
\begin{equation}
\label{eq:rzzprompt}
\rzz = \fprompt \cdot \rzzp + (1 - \fprompt) \cdot \rzznp,
\end{equation}
where $\fprompt$ is the fraction of prompt \Dstar mesons in the raw yields. In order to evaluate this fraction, a data-driven method based on the sampling of the raw yield $Y_\mathrm{i}$ at different values of the BDT-output score related to the probability of being a non-prompt \Dstar meson was adopted~\cite{ALICE:2021mgk}. The BDT score related to the non-prompt \Dstar meson probability was adopted rather than the one related to the prompt \Dstar probability owing its more powerful discrimination of the two contributions.
The \fprompt parameter can be computed by solving a system of equations that relate raw yields to the corrected yields of prompt (\Np) and non-prompt (\Nnp) \Dstar mesons by the \AccEff factors. Each equation, obtained for a set of BDT selections $i$, can be expressed as
\begin{equation}
   \effP{i}\cdot \Np +  \effNP{i}\cdot \Nnp - \rawY{i} = \delta_i,
\label{eq:eq_set}
\end{equation}
where $\delta_i$ is the residual originating from the uncertainties on \rawY{i}, \effNP{i}, and \effP{i}. The sets of selections were defined sampling the BDT-output score related to the probability of a candidate to be a non-prompt \Dstar meson monotonically, to obtain $n$ selections ordered in such a way that the $i^\mathrm{th}$ selected sample was completely included in the $(i-1)^\mathrm{th}$ one. The system of equations is then solved via a $\chi^2$ minimisation procedure, which provides as output the corrected yields of prompt (\Np) and non-prompt (\Nnp) \Dstar mesons and their covariance matrix. Finally, \fprompt was determined as 
\begin{equation}
    f_{\mathrm{prompt},j} = \frac{\effP{j} \cdot \Np}{\effP{j}\cdot\Np+\effNP{j}\cdot\Nnp}.
\end{equation}
where $j$ denotes the set of selections used for the extraction of the \rzz parameter, which is different from those adopted in the fraction evaluation.
The uncertainty of $f_{\mathrm{prompt},j}$ is computed considering the covariance matrix of \Np and \Nnp obtained in the $\chi^2$-minimisation procedure.

\begin{figure}[!t]
\begin{center}
\includegraphics[width=1.\textwidth]{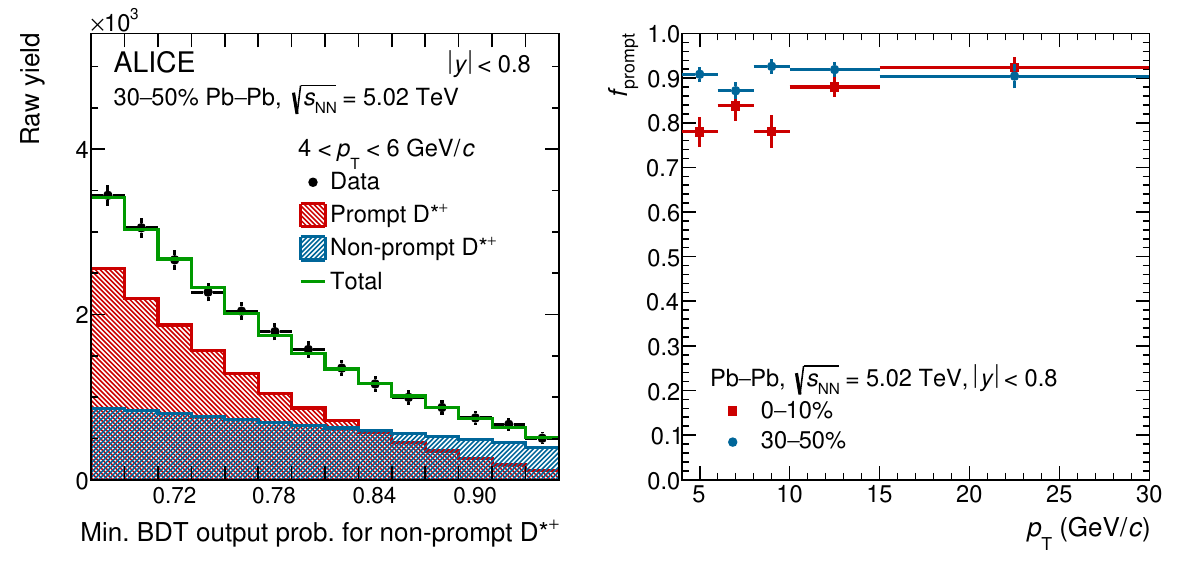}
\caption{Left: raw yield as a function of the BDT selection and extracted contributions from prompt and non-prompt \Dstar mesons with $4<\pt<6~\GeV/c$ and $|y|<0.8$ in the 30--50\% centrality class. Right: measured fraction of prompt \Dstar mesons as a function of \pt in the 0--10\% and 30--50\% centrality classes. Only the statistical uncertainties are reported.}
\label{fig:cutvar}
\end{center}
\end{figure}

The left panel of Fig.~\ref{fig:cutvar} shows an example of a raw yield extracted as a function of the BDT selection employed in the minimisation procedure for \Dstar mesons with $4<\pt<6~\GeV/c$ in midcentral Pb--Pb collisions. The leftmost data point of the distribution is the raw yield corresponding to the loosest selection on the BDT output related to the candidate probability of being a non-prompt \Dstar meson, while the rightmost one corresponds to the strictest selection, which is expected to preferentially select non-prompt \Dstar mesons. The prompt and non-prompt components are represented by the red and blue filled histograms, respectively, while their sum is depicted by the green histogram. The \fprompt factor obtained in the 0--10\% and 30--50\% centrality classes for the selections adopted for the \rzz extraction is reported in the right panel of the same figure. The fraction was evaluated for the total rapidity range $|y|<0.8$, considering the limited size of the data sample and that no rapidity dependence is expected within the analysed range.

Finally, to obtain the \rzzp parameter from Eq.~\ref{eq:rzzprompt} an assumption on the spin alignment of non-prompt \Dstar mesons was necessary. The \rzz parameter of non-prompt \Dstar mesons was measured in pp collisions and found to be larger than $1/3$ with respect to the helicity axis, i.e. with respect to the direction of the \Dstar momentum in the laboratory frame~\cite{ALICE:2022byg}. This result was understood as a consequence of the helicity conservation in scalar beauty-hadron decays. The measured \rzz parameter of non-prompt \Dstar mesons was found also to be well described by \pythia + \evtgen simulations.
Given that the \Dstar momentum (i.e. the helicity axis) in heavy-ion collisions is correlated with the direction of the reaction plane owing to the azimuthal anisotropies, due to the positive second-harmonic flow coefficient $v_2$ (elliptic flow), non-prompt \Dstar mesons are expected to exhibit a \rzz larger than $1/3$ with respect to the second-harmonic event plane $\psi_2$. This implies a \rzz lower than $1/3$ in the case of the quantisation axis considered in this analysis, which is orthogonal to $\psi_2$ and the beam axis. Following the study published in Ref.~\cite{ALICE:2022byg}, \rzznp was then computed by performing \pythia + \evtgen simulations, in which an elliptic flow modulation was included. In particular, $v_2 = 0.05$ was used for the elliptic flow of non-prompt \Dstar mesons, based on the recent publications of the measurement of non-prompt \Dzero meson $v_2$ by the ALICE~\cite{ALICE:2023gjj} and CMS~\cite{CMS:2022vfn} Collaborations.

Considering the high prompt fraction (right panel of Fig.~\ref{fig:cutvar}) and the relatively small $v_2$ of non-prompt \Dstar mesons, the procedure described above led to a maximum correction of about $0.01$ in absolute value to the measured \rzz to obtain the one of prompt \Dstar mesons.

\section{Systematic uncertainties}
\label{sec:syst}

The main sources of systematic uncertainties considered for the measurement of the \rzz parameter of prompt \Dstar mesons are i) the signal extraction, ii) the track reconstruction and selection efficiencies, iii) the BDT selection efficiency, and iv) the subtraction of the non-prompt contribution. The estimated values of the systematic uncertainties are summarised in Table~\ref{tab:sysunc} for representative \pt intervals, together with the total systematic uncertainty obtained from the sum in quadrature of the different contributions. Since no significant difference in the systematic uncertainties evaluated in the two rapidity intervals was found, the same values were assigned to the measurements performed in the $|y| < 0.3$ and $0.3 < |y| < 0.8$ intervals.

The systematic uncertainty of the raw-yield extraction was evaluated by repeating the fit of the invariant-mass distribution varying the lower and upper limits of the fit range, leaving the Gaussian width free in the fits, and using an alternative functional form to describe the combinatorial background, namely $p_0(\Delta M - M_\pi)^{p_1}$, where $p_0$ and $p_1$ are free parameters. The values of systematic uncertainties range from 13\% to 17\% for central collisions, and 3\% to 11\% for midcentral collisions depending on \pt.

The systematic uncertainty associated to the track selection and reconstruction efficiency was evaluated by varying the track-quality selection criteria and comparing the different \rzz parameters obtained. In particular, different requirements of number of space points for the track-reconstruction in the ITS and TPC were tested. The systematic uncertainty ranges between 3\% and 8\% for central collisions and 2\% and 4\% for midcentral collisions, depending on \pt.

The systematic uncertainty associated to the determination of the selection efficiency, arising from possible imperfections in the description of the decay topology or the detector resolution in the simulation, was estimated by using alternative sets of selections on the BDT-output score related to the probability of being background and repeating the \rzz measurement. A systematic uncertainty ranging between 3\% and 7\% was assigned for central collisions, while an uncertainty between 3\% and 5\% for midcentral collisions was assigned.

\begin{table}[!htb]
\caption{Summary of the relative systematic uncertainties on the \rzz parameter of prompt \Dstar mesons for representative \pt intervals in the 0--10\% and 30--50\% centrality classes.}
\centering
\renewcommand*{\arraystretch}{1.2}
\begin{tabular}[t]{l|>{\centering}p{0.10\linewidth}>{\centering}p{0.10\linewidth}|>{\centering}p{0.10\linewidth}>{\centering\arraybackslash}p{0.10\linewidth}}
\toprule
 & \multicolumn{2}{c|}{0--10\%} & \multicolumn{2}{c}{30--50\%} \\
\midrule
$\pt~(\GeV/c)$ & $4\mbox{--}6$ & $15\mbox{--}30$ & $4\mbox{--}6$ & $15\mbox{--}30$ \\
\midrule
Signal yield & 17\% & 13\% & 5\% & 3\% \\
Tracking efficiency & 6\% & 5\% & 4\% & 2\% \\
BDT efficiency & 5\% & 3\% & 5\% & 3\% \\
Non-prompt subtraction & negl. & negl. & negl. & negl. \\
\midrule
Total & 19\% & 14\% & 8\% & 5\% \\
\bottomrule
\end{tabular}
\label{tab:sysunc}	
\end{table}
The systematic uncertainty related to the subtraction of the non-prompt component has two contributions. The first one is related to the determination of the prompt fraction in the raw yield. Similarly to the previous source of systematic uncertainty, this is mainly related to the description of the topological variables in the MC simulations used to determine the selection efficiencies of prompt and non-prompt \Dstar mesons. Therefore, stricter and looser selections on the BDT score related to be a non-prompt \Dstar meson, as well as different combinations of selections adopted to define the system of equations described in Eq.~\ref{eq:eq_set}, were tested. The second source of uncertainty instead originates from the assumption on the elliptic flow of non-prompt \Dstar mesons. In this case, a value of $v_2 = 0.10$ was used alternatively to the default one of $v_2 = 0.05$. Both the effect of the \fprompt and $v_2$ variations were found to have a negligible effect on the \rzz parameter for both centrality classes.

\section{Results}
\label{sec:res}

\begin{figure}[!tb]
\begin{center}
\includegraphics[width=1.\textwidth]{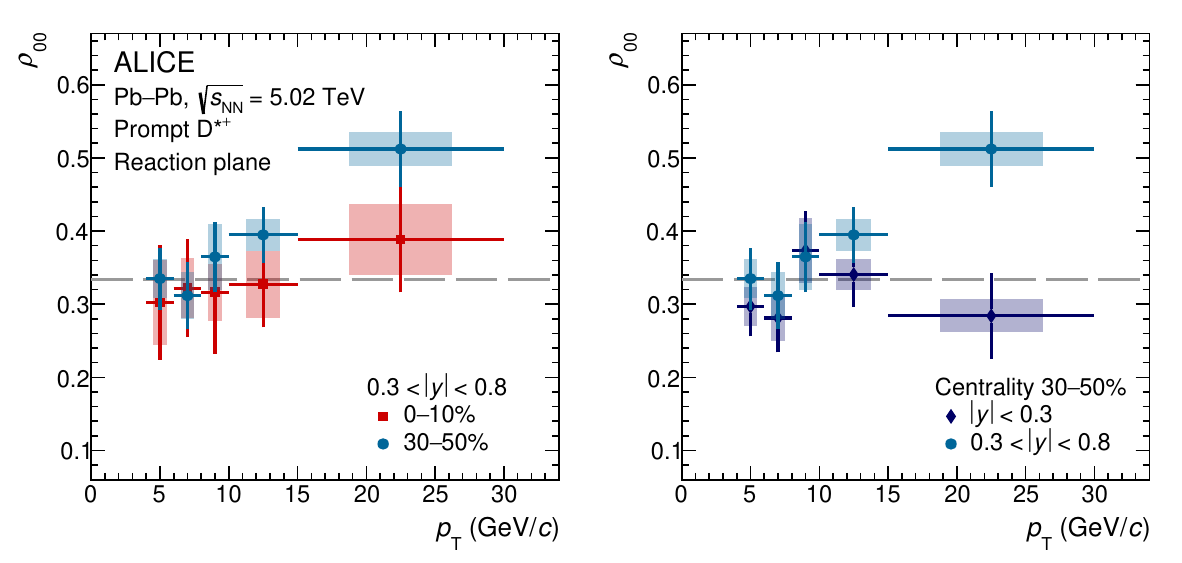}
\caption{Left: the spin density matrix element \rzz of prompt \Dstar mesons as a function of \pt in the rapidity interval $0.3 < |y| < 0.8$ for central (0--10\%) and midcentral (30--50\%) Pb--Pb collisions at $\snn=5.02~\TeV$. Right: \rzz of prompt \Dstar mesons as a function of \pt in the two rapidity intervals for midcentral (30--50\%) Pb--Pb collisions at $\snn=5.02~\TeV$. The measurements are carried out with respect to the direction perpendicular to the reaction plane. Statistical and systematic uncertainties are represented by bars and boxes, respectively.} 
\label{fig:result} 
\end{center}
\end{figure}

The left panel of Fig.~\ref{fig:result} shows the \rzz parameter measured with respect to the direction orthogonal to the reaction plane for prompt \Dstar mesons with $0.3<|y|<0.8$ as a function of \pt in central (0--10\%) and midcentral (30--50\%) Pb--Pb collisions at $\snn=5.02~\TeV$. Vertical bars represent statistical uncertainties, while the filled boxes represent systematic uncertainties. The data points for central events are compatible with $1/3$ in the full \pt range. A hint of increasing trend with \pt is instead found for midcentral events, with a maximum deviation from $1/3$ in the $15<\pt<30~\GeV/c$ interval of about $3.1\sigma$, where $\sigma$ is the sum in quadrature of statistical and systematic uncertainties. However, no significant deviation from $1/3$ is observed for \Dstar mesons in the $|y|<0.3$ interval even at high \pt, as reported in the right panel of Fig.~\ref{fig:result}.

The intriguing possible \pt and rapidity dependence of \Dstar spin alignment measured at the LHC energies may be explained by a polarisation of early produced high-\pt charm quarks induced by the early magnetic field, which decreases slower in time as rapidity increases~\cite{Das:2016cwd, Dubla:2020bdz}. 
In the thermal limit, there are currently no theory calculations based on vorticity and magnetic field that could lead to the observed values of vector-meson spin alignment. However, it is worth to note that the measured spin alignment of \Dstar meson seems to be larger at very high \pt, where the charm quarks are less likely to be thermalised. 
In fact, a recent theory study based on rotational Brownian motion of heavy quarks in a QCD medium suggests that heavy quarks with higher \pt can retain a sizable polarisation originating from the initial magnetic field due to their shorter interaction time with the medium constituents~\cite{Dey:2025ail}. In particular, a significant enhancement of charm-quark polarisation with \pt is predicted, with the magnitude of charm-quark polarisation depending on the spin-relaxation time of charm quarks. 
Moreover, the high value of spin alignment could have a different physical origin, e.g. local polarisation~\cite{local1, local2}, fluctuation of quark polarisation in turbulent colour fields~\cite{Muller:2021hpe}, local spin correlation in the glasma phase~\cite{Kumar:2023ghs}, \Ph meson fields~\cite{Sheng:2022ffb,Sheng:2023urn}, modification of vector-meson spectral function in thermal background~\cite{Sheng:2024kgg}. Theory calculations based on \Ph meson field~\cite{Sheng:2022ffb,Sheng:2023urn} predict a non-trivial transverse momentum and rapidity dependence of the spin alignment. In particular, the \pt dependence of the spin alignment is correlated with the elliptic flow, therefore a significant positive spin alignment ($\rzz > 1/3$) is expected at relatively forward rapidity. Similar rapidity and \pt dependence is also predicted by the theory calculation based on the holographic spin alignment~\cite{Sheng:2024kgg} that incorporates the modification of the vector-meson spectral function for different spin modes in the presence of thermal background. It is important to notice that these calculations are valid only in the limit of a static medium and neglect the hydrodynamic expansion of the QGP. However, corrections due to the hydrodynamic expansion of the medium are expected to not have a significant impact for high-momentum partons, as discussed in Ref.~\cite{Xu:2015bbz}.
Although the possible rapidity dependence measured for the \Dstar spin alignment is qualitatively similar to these theory calculations, these predictions are only available for flavourless vector mesons (\Ph and \Jpsi).
To compare \Dstar spin alignment measurements with meson field theory calculations~\cite{Sheng:2022ffb,Sheng:2023urn},
it is necessary to incorporate both the \Dstar field and the fragmentation hadronisation mechanism into these models. This is because current calculations assume coalescence as the primary hadronisation process, while fragmentation is expected to dominate in the high-$\pt$ region~\cite{Plumari:2017ntm,Beraudo:2022dpz}. Additionally, theoretical predictions~\cite{Liang:2004xn} suggest that the fragmentation of polarised charm quarks could result in a \rzz value larger than 1/3.
This measurement will motivate new theoretical developments in the charm-hadron sector in order to understand the possible underlying physics mechanism of charm-hadron polarisation.

\begin{figure}[!t]
\begin{center}
\includegraphics[width=0.6\textwidth]{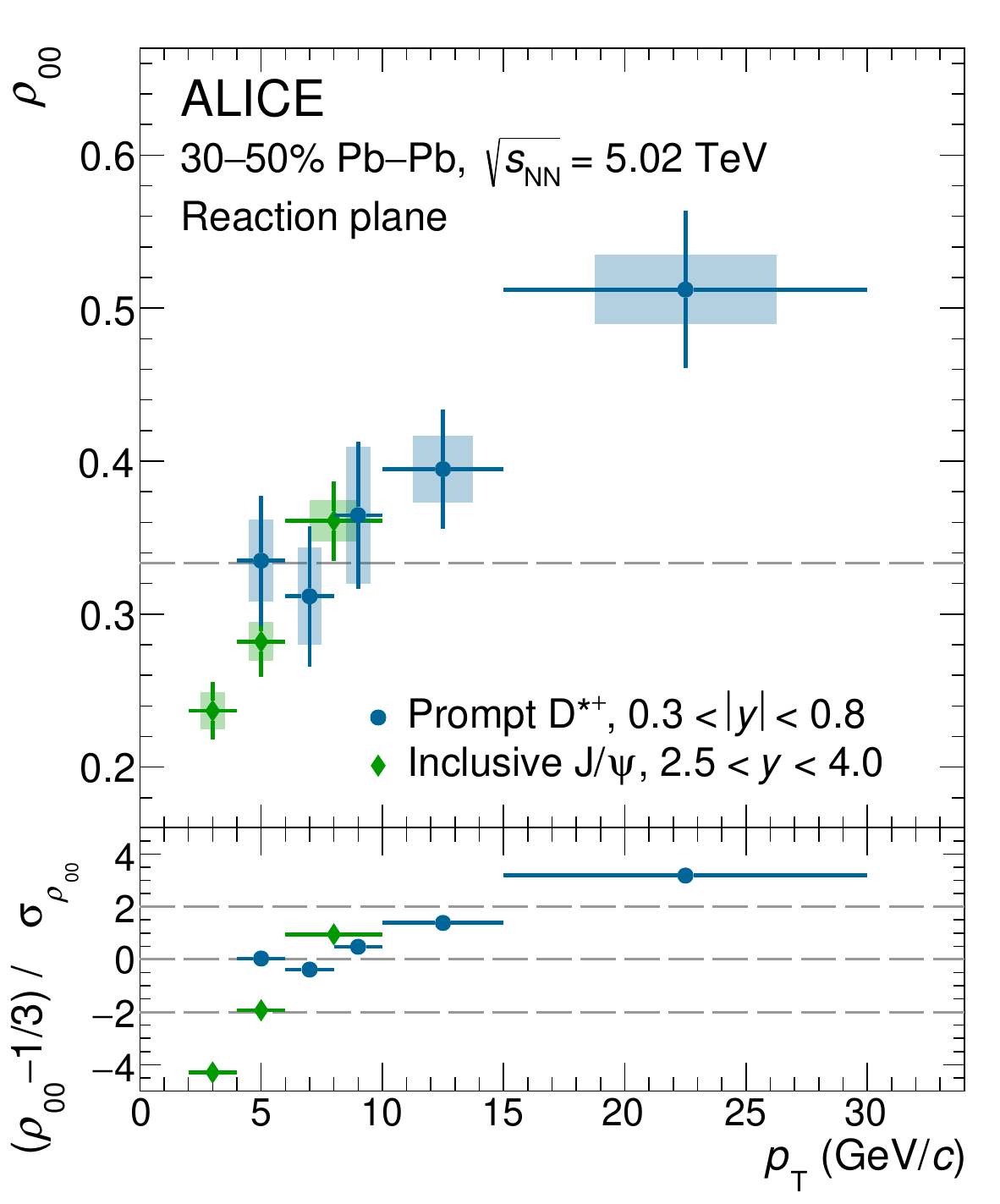}
\caption{The spin density matrix element \rzz of prompt \Dstar mesons as a function of \pt in the rapidity interval $0.3 < |y| < 0.8$ compared to that of inclusive \Jpsi mesons in the rapidity interval $2.5<y<4$ measured in Pb--Pb collisions at $\snn=5.02~\TeV$ in the 30--50\% centrality class. The bottom panel shows the deviation of the measurements from the null hypothesis ($\rzz = 1/3$) in units of the total uncertainties $\sigma_{\rzz}$.}
\label{fig:result_vsjpsi} 
\end{center}
\end{figure}

Figure~\ref{fig:result_vsjpsi} shows the comparison of the \rzz parameter of prompt \Dstar mesons in the rapidity interval $0.3<|y|<0.8$ compared with the one of inclusive \Jpsi mesons measured in the $2.5<y<4$ rapidity interval in Pb--Pb collisions at $\snn=5.02~\TeV$ for the 30--50\% centrality class~\cite{ALICE:2022dyy}. Measurements of quarkonia polarisation are typically reported in terms of the \lt, \lp, and \ltp parameters, which are extracted from the two-dimensional angular distribution of the two leptons (\ee or \mumu) emerging from their decays as reported in Refs.~\cite{ALICE:2011gej,ALICE:2018crw,LHCb:2013izl,CMS:2013gbz,ALICE:2020iev,ALICE:2022dyy}. In particular, the \lt parameter is related the \rzz parameter via the following relation
\begin{equation}
   \rzz = \frac{1-\lt}{3+\lt},
\end{equation}
implying $\rzz>1/3$ if $\lt<0$ and viceversa. The bottom panel of the same figure shows the deviation of \rzz from $\rzz=1/3$ in units of the total uncertainties $\sigma$. An evidence of $\rzz<1/3$ is found for inclusive \Jpsi mesons at forward rapidity with $\pt<4~\GeV/c$ in midcentral Pb--Pb collisions which is qualitatively in agreement with a recent theory calculation based on holographic spin alignment~\cite{Sheng:2024kgg}. The measured \rzz of prompt \Dstar mesons in the rapidity range $0.3<|y|<0.8$ and the inclusive \Jpsi at forward rapidity ($2.5<y<4$) are compatible within uncertainties in the overlapping \pt region ($4<\pt<10~\GeV/c$) and seem to feature a common increasing trend with increasing \pt. However, in order to draw firm conclusions about a possible common dependence with \pt, proper theoretical guidance is needed.

\section{Summary}
\label{sec:sum}
In summary, the first measurement of prompt \Dstar-meson spin alignment with respect to the direction orthogonal to the reaction plane at midrapidity in central (0--10\%) and midcentral Pb--Pb collisions at $\snn = 5.02~\TeV$ has been presented. An evidence of $\rzz>1/3$ for $\pt > 15~\GeV/c$ in the $0.3<|y|<0.8$ range has been found for midcentral (30--50\%) Pb--Pb collisions.
The measured spin alignment seems to be more prominent at very high-\pt where charm quarks are expected to be produced much earlier and less likely to be thermalised. In order to understand any possible effect of early magnetic fields created in heavy-ion collisions on charm-quark polarisation, dedicated theory predictions of charm-quark polarisation and charm-hadron spin alignment are needed. From the experimental side, significant precision improvements for charm-hadron measurements are expected from the large datasets that are being collected by the ALICE Collaboration during the LHC Run 3 data-taking period, thanks to the upgrade of the experimental apparatus~\cite{ALICE:2023udb}. This will also allow for the extension of polarisation and spin alignment measurements of other charm-hadron species, such as \Lc baryons.

\newenvironment{acknowledgement}{\relax}{\relax}
\begin{acknowledgement}
\section*{Acknowledgements}
% Version: 2025-03-18

The ALICE Collaboration would like to thank all its engineers and technicians for their invaluable contributions to the construction of the experiment and the CERN accelerator teams for the outstanding performance of the LHC complex.
The ALICE Collaboration gratefully acknowledges the resources and support provided by all Grid centres and the Worldwide LHC Computing Grid (WLCG) collaboration.
The ALICE Collaboration acknowledges the following funding agencies for their support in building and running the ALICE detector:
A. I. Alikhanyan National Science Laboratory (Yerevan Physics Institute) Foundation (ANSL), State Committee of Science and World Federation of Scientists (WFS), Armenia;
Austrian Academy of Sciences, Austrian Science Fund (FWF): [M 2467-N36] and Nationalstiftung f\"{u}r Forschung, Technologie und Entwicklung, Austria;
Ministry of Communications and High Technologies, National Nuclear Research Center, Azerbaijan;
Conselho Nacional de Desenvolvimento Cient\'{\i}fico e Tecnol\'{o}gico (CNPq), Financiadora de Estudos e Projetos (Finep), Funda\c{c}\~{a}o de Amparo \`{a} Pesquisa do Estado de S\~{a}o Paulo (FAPESP) and Universidade Federal do Rio Grande do Sul (UFRGS), Brazil;
Bulgarian Ministry of Education and Science, within the National Roadmap for Research Infrastructures 2020-2027 (object CERN), Bulgaria;
Ministry of Education of China (MOEC) , Ministry of Science \& Technology of China (MSTC) and National Natural Science Foundation of China (NSFC), China;
Ministry of Science and Education and Croatian Science Foundation, Croatia;
Centro de Aplicaciones Tecnol\'{o}gicas y Desarrollo Nuclear (CEADEN), Cubaenerg\'{\i}a, Cuba;
Ministry of Education, Youth and Sports of the Czech Republic, Czech Republic;
The Danish Council for Independent Research | Natural Sciences, the VILLUM FONDEN and Danish National Research Foundation (DNRF), Denmark;
Helsinki Institute of Physics (HIP), Finland;
Commissariat \`{a} l'Energie Atomique (CEA) and Institut National de Physique Nucl\'{e}aire et de Physique des Particules (IN2P3) and Centre National de la Recherche Scientifique (CNRS), France;
Bundesministerium f\"{u}r Bildung und Forschung (BMBF) and GSI Helmholtzzentrum f\"{u}r Schwerionenforschung GmbH, Germany;
General Secretariat for Research and Technology, Ministry of Education, Research and Religions, Greece;
National Research, Development and Innovation Office, Hungary;
Department of Atomic Energy Government of India (DAE), Department of Science and Technology, Government of India (DST), University Grants Commission, Government of India (UGC) and Council of Scientific and Industrial Research (CSIR), India;
National Research and Innovation Agency - BRIN, Indonesia;
Istituto Nazionale di Fisica Nucleare (INFN), Italy;
Japanese Ministry of Education, Culture, Sports, Science and Technology (MEXT) and Japan Society for the Promotion of Science (JSPS) KAKENHI, Japan;
Consejo Nacional de Ciencia (CONACYT) y Tecnolog\'{i}a, through Fondo de Cooperaci\'{o}n Internacional en Ciencia y Tecnolog\'{i}a (FONCICYT) and Direcci\'{o}n General de Asuntos del Personal Academico (DGAPA), Mexico;
Nederlandse Organisatie voor Wetenschappelijk Onderzoek (NWO), Netherlands;
The Research Council of Norway, Norway;
Pontificia Universidad Cat\'{o}lica del Per\'{u}, Peru;
Ministry of Science and Higher Education, National Science Centre and WUT ID-UB, Poland;
Korea Institute of Science and Technology Information and National Research Foundation of Korea (NRF), Republic of Korea;
Ministry of Education and Scientific Research, Institute of Atomic Physics, Ministry of Research and Innovation and Institute of Atomic Physics and Universitatea Nationala de Stiinta si Tehnologie Politehnica Bucuresti, Romania;
Ministerstvo skolstva, vyskumu, vyvoja a mladeze SR, Slovakia;
National Research Foundation of South Africa, South Africa;
Swedish Research Council (VR) and Knut \& Alice Wallenberg Foundation (KAW), Sweden;
European Organization for Nuclear Research, Switzerland;
Suranaree University of Technology (SUT), National Science and Technology Development Agency (NSTDA) and National Science, Research and Innovation Fund (NSRF via PMU-B B05F650021), Thailand;
Turkish Energy, Nuclear and Mineral Research Agency (TENMAK), Turkey;
National Academy of  Sciences of Ukraine, Ukraine;
Science and Technology Facilities Council (STFC), United Kingdom;
National Science Foundation of the United States of America (NSF) and United States Department of Energy, Office of Nuclear Physics (DOE NP), United States of America.
In addition, individual groups or members have received support from:
Czech Science Foundation (grant no. 23-07499S), Czech Republic;
FORTE project, reg.\ no.\ CZ.02.01.01/00/22\_008/0004632, Czech Republic, co-funded by the European Union, Czech Republic;
European Research Council (grant no. 950692), European Union;
Deutsche Forschungs Gemeinschaft (DFG, German Research Foundation) ``Neutrinos and Dark Matter in Astro- and Particle Physics'' (grant no. SFB 1258), Germany;
ICSC - National Research Center for High Performance Computing, Big Data and Quantum Computing and FAIR - Future Artificial Intelligence Research, funded by the NextGenerationEU program (Italy).
 
%%%%%%% done by webmaster team
\end{acknowledgement}
\bibliographystyle{utphys}
\bibliography{dstar_spin_alignment}
    
\newpage
\appendix
\section{The ALICE Collaboration}
\label{app:collab}
% ALICE Collaboration author list for 2025-03-18
\begin{flushleft} 
\small

S.~Acharya\,\orcidlink{0000-0002-9213-5329}\,$^{\rm 50}$, 
A.~Agarwal$^{\rm 133}$, 
G.~Aglieri Rinella\,\orcidlink{0000-0002-9611-3696}\,$^{\rm 32}$, 
L.~Aglietta\,\orcidlink{0009-0003-0763-6802}\,$^{\rm 24}$, 
M.~Agnello\,\orcidlink{0000-0002-0760-5075}\,$^{\rm 29}$, 
N.~Agrawal\,\orcidlink{0000-0003-0348-9836}\,$^{\rm 25}$, 
Z.~Ahammed\,\orcidlink{0000-0001-5241-7412}\,$^{\rm 133}$, 
S.~Ahmad\,\orcidlink{0000-0003-0497-5705}\,$^{\rm 15}$, 
S.U.~Ahn\,\orcidlink{0000-0001-8847-489X}\,$^{\rm 71}$, 
I.~Ahuja\,\orcidlink{0000-0002-4417-1392}\,$^{\rm 36}$
% ...

\section*{Affiliation Notes}

$^{\rm I}$ Deceased\\
$^{\rm II}$ Also at: Max-Planck-Institut fur Physik, Munich, Germany\\
$^{\rm III}$ Also at: Italian National Agency for New Technologies, Energy and Sustainable Economic Development (ENEA), Bologna, Italy\\
$^{\rm IV}$ Also at: Instituto de Fisica da Universidade de Sao Paulo\\
$^{\rm V}$ Also at: Dipartimento DET del Politecnico di Torino, Turin, Italy\\
$^{\rm VI}$ Also at: Department of Applied Physics, Aligarh Muslim University, Aligarh, India\\
$^{\rm VII}$ Also at: Institute of Theoretical Physics, University of Wroclaw, Poland\\
$^{\rm VIII}$ Also at: Facultad de Ciencias, Universidad Nacional Aut\'{o}noma de M\'{e}xico, Mexico City, Mexico\\

\section*{Collaboration Institutes}

$^{1}$ A.I. Alikhanyan National Science Laboratory (Yerevan Physics Institute) Foundation, Yerevan, Armenia\\
$^{2}$ AGH University of Krakow, Cracow, Poland\\
$^{3}$ Bogolyubov Institute for Theoretical Physics, National Academy of Sciences of Ukraine, Kiev, Ukraine\\
$^{4}$ Bose Institute, Department of Physics  and Centre for Astroparticle Physics and Space Science (CAPSS), Kolkata, India\\
$^{5}$ California Polytechnic State University, San Luis Obispo, California, United States\\
$^{6}$ Central China Normal University, Wuhan, China\\
$^{7}$ Centro de Aplicaciones Tecnol\'{o}gicas y Desarrollo Nuclear (CEADEN), Havana, Cuba\\
$^{8}$ Centro de Investigaci\'{o}n y de Estudios Avanzados (CINVESTAV), Mexico City and M\'{e}rida, Mexico\\
$^{9}$ Chicago State University, Chicago, Illinois, United States\\
$^{10}$ China Nuclear Data Center, China Institute of Atomic Energy, Beijing, China\\
$^{11}$ China University of Geosciences, Wuhan, China\\
$^{12}$ Chungbuk National University, Cheongju, Republic of Korea\\
$^{13}$ Comenius University Bratislava, Faculty of Mathematics, Physics and Informatics, Bratislava, Slovak Republic\\
$^{14}$ Creighton University, Omaha, Nebraska, United States\\
$^{15}$ Department of Physics, Aligarh Muslim University, Aligarh, India\\
$^{16}$ Department of Physics, Pusan National University, Pusan, Republic of Korea\\
$^{17}$ Department of Physics, Sejong University, Seoul, Republic of Korea\\
$^{18}$ Department of Physics, University of California, Berkeley, California, United States\\
$^{19}$ Department of Physics, University of Oslo, Oslo, Norway\\
$^{20}$ Department of Physics and Technology, University of Bergen, Bergen, Norway\\
$^{21}$ Dipartimento di Fisica, Universit\`{a} di Pavia, Pavia, Italy\\
$^{22}$ Dipartimento di Fisica dell'Universit\`{a} and Sezione INFN, Cagliari, Italy\\
$^{23}$ Dipartimento di Fisica dell'Universit\`{a} and Sezione INFN, Trieste, Italy\\
$^{24}$ Dipartimento di Fisica dell'Universit\`{a} and Sezione INFN, Turin, Italy\\
$^{25}$ Dipartimento di Fisica e Astronomia dell'Universit\`{a} and Sezione INFN, Bologna, Italy\\
$^{26}$ Dipartimento di Fisica e Astronomia dell'Universit\`{a} and Sezione INFN, Catania, Italy\\
$^{27}$ Dipartimento di Fisica e Astronomia dell'Universit\`{a} and Sezione INFN, Padova, Italy\\
$^{28}$ Dipartimento di Fisica `E.R.~Caianiello' dell'Universit\`{a} and Gruppo Collegato INFN, Salerno, Italy\\
$^{29}$ Dipartimento DISAT del Politecnico and Sezione INFN, Turin, Italy\\
$^{30}$ Dipartimento di Scienze MIFT, Universit\`{a} di Messina, Messina, Italy\\
$^{31}$ Dipartimento Interateneo di Fisica `M.~Merlin' and Sezione INFN, Bari, Italy\\
$^{32}$ European Organization for Nuclear Research (CERN), Geneva, Switzerland\\
$^{33}$ Faculty of Electrical Engineering, Mechanical Engineering and Naval Architecture, University of Split, Split, Croatia\\
$^{34}$ Faculty of Nuclear Sciences and Physical Engineering, Czech Technical University in Prague, Prague, Czech Republic\\
$^{35}$ Faculty of Physics, Sofia University, Sofia, Bulgaria\\
$^{36}$ Faculty of Science, P.J.~\v{S}af\'{a}rik University, Ko\v{s}ice, Slovak Republic\\
$^{37}$ Faculty of Technology, Environmental and Social Sciences, Bergen, Norway\\
$^{38}$ Frankfurt Institute for Advanced Studies, Johann Wolfgang Goethe-Universit\"{a}t Frankfurt, Frankfurt, Germany\\
$^{39}$ Fudan University, Shanghai, China\\
$^{40}$ Gangneung-Wonju National University, Gangneung, Republic of Korea\\
$^{41}$ Gauhati University, Department of Physics, Guwahati, India\\
$^{42}$ Helmholtz-Institut f\"{u}r Strahlen- und Kernphysik, Rheinische Friedrich-Wilhelms-Universit\"{a}t Bonn, Bonn, Germany\\
$^{43}$ Helsinki Institute of Physics (HIP), Helsinki, Finland\\
$^{44}$ High Energy Physics Group,  Universidad Aut\'{o}noma de Puebla, Puebla, Mexico\\
$^{45}$ Horia Hulubei National Institute of Physics and Nuclear Engineering, Bucharest, Romania\\
$^{46}$ HUN-REN Wigner Research Centre for Physics, Budapest, Hungary\\
$^{47}$ Indian Institute of Technology Bombay (IIT), Mumbai, India\\
$^{48}$ Indian Institute of Technology Indore, Indore, India\\
$^{49}$ INFN, Laboratori Nazionali di Frascati, Frascati, Italy\\
$^{50}$ INFN, Sezione di Bari, Bari, Italy\\
$^{51}$ INFN, Sezione di Bologna, Bologna, Italy\\
$^{52}$ INFN, Sezione di Cagliari, Cagliari, Italy\\
$^{53}$ INFN, Sezione di Catania, Catania, Italy\\
$^{54}$ INFN, Sezione di Padova, Padova, Italy\\
$^{55}$ INFN, Sezione di Pavia, Pavia, Italy\\
$^{56}$ INFN, Sezione di Torino, Turin, Italy\\
$^{57}$ INFN, Sezione di Trieste, Trieste, Italy\\
$^{58}$ Inha University, Incheon, Republic of Korea\\
$^{59}$ Institute for Gravitational and Subatomic Physics (GRASP), Utrecht University/Nikhef, Utrecht, Netherlands\\
$^{60}$ Institute of Experimental Physics, Slovak Academy of Sciences, Ko\v{s}ice, Slovak Republic\\
$^{61}$ Institute of Physics, Homi Bhabha National Institute, Bhubaneswar, India\\
$^{62}$ Institute of Physics of the Czech Academy of Sciences, Prague, Czech Republic\\
$^{63}$ Institute of Space Science (ISS), Bucharest, Romania\\
$^{64}$ Institut f\"{u}r Kernphysik, Johann Wolfgang Goethe-Universit\"{a}t Frankfurt, Frankfurt, Germany\\
$^{65}$ Instituto de Ciencias Nucleares, Universidad Nacional Aut\'{o}noma de M\'{e}xico, Mexico City, Mexico\\
$^{66}$ Instituto de F\'{i}sica, Universidade Federal do Rio Grande do Sul (UFRGS), Porto Alegre, Brazil\\
$^{67}$ Instituto de F\'{\i}sica, Universidad Nacional Aut\'{o}noma de M\'{e}xico, Mexico City, Mexico\\
$^{68}$ iThemba LABS, National Research Foundation, Somerset West, South Africa\\
$^{69}$ Jeonbuk National University, Jeonju, Republic of Korea\\
$^{70}$ Johann-Wolfgang-Goethe Universit\"{a}t Frankfurt Institut f\"{u}r Informatik, Fachbereich Informatik und Mathematik, Frankfurt, Germany\\
$^{71}$ Korea Institute of Science and Technology Information, Daejeon, Republic of Korea\\
$^{72}$ Laboratoire de Physique Subatomique et de Cosmologie, Universit\'{e} Grenoble-Alpes, CNRS-IN2P3, Grenoble, France\\
$^{73}$ Lawrence Berkeley National Laboratory, Berkeley, California, United States\\
$^{74}$ Lund University Department of Physics, Division of Particle Physics, Lund, Sweden\\
$^{75}$ Nagasaki Institute of Applied Science, Nagasaki, Japan\\
$^{76}$ Nara Women{'}s University (NWU), Nara, Japan\\
$^{77}$ National and Kapodistrian University of Athens, School of Science, Department of Physics , Athens, Greece\\
$^{78}$ National Centre for Nuclear Research, Warsaw, Poland\\
$^{79}$ National Institute of Science Education and Research, Homi Bhabha National Institute, Jatni, India\\
$^{80}$ National Nuclear Research Center, Baku, Azerbaijan\\
$^{81}$ National Research and Innovation Agency - BRIN, Jakarta, Indonesia\\
$^{82}$ Niels Bohr Institute, University of Copenhagen, Copenhagen, Denmark\\
$^{83}$ Nikhef, National institute for subatomic physics, Amsterdam, Netherlands\\
$^{84}$ Nuclear Physics Group, STFC Daresbury Laboratory, Daresbury, United Kingdom\\
$^{85}$ Nuclear Physics Institute of the Czech Academy of Sciences, Husinec-\v{R}e\v{z}, Czech Republic\\
$^{86}$ Oak Ridge National Laboratory, Oak Ridge, Tennessee, United States\\
$^{87}$ Ohio State University, Columbus, Ohio, United States\\
$^{88}$ Physics department, Faculty of science, University of Zagreb, Zagreb, Croatia\\
$^{89}$ Physics Department, Panjab University, Chandigarh, India\\
$^{90}$ Physics Department, University of Jammu, Jammu, India\\
$^{91}$ Physics Program and International Institute for Sustainability with Knotted Chiral Meta Matter (WPI-SKCM$^{2}$), Hiroshima University, Hiroshima, Japan\\
$^{92}$ Physikalisches Institut, Eberhard-Karls-Universit\"{a}t T\"{u}bingen, T\"{u}bingen, Germany\\
$^{93}$ Physikalisches Institut, Ruprecht-Karls-Universit\"{a}t Heidelberg, Heidelberg, Germany\\
$^{94}$ Physik Department, Technische Universit\"{a}t M\"{u}nchen, Munich, Germany\\
$^{95}$ Politecnico di Bari and Sezione INFN, Bari, Italy\\
$^{96}$ Research Division and ExtreMe Matter Institute EMMI, GSI Helmholtzzentrum f\"ur Schwerionenforschung GmbH, Darmstadt, Germany\\
$^{97}$ Saga University, Saga, Japan\\
$^{98}$ Saha Institute of Nuclear Physics, Homi Bhabha National Institute, Kolkata, India\\
$^{99}$ School of Physics and Astronomy, University of Birmingham, Birmingham, United Kingdom\\
$^{100}$ Secci\'{o}n F\'{\i}sica, Departamento de Ciencias, Pontificia Universidad Cat\'{o}lica del Per\'{u}, Lima, Peru\\
$^{101}$ Stefan Meyer Institut f\"{u}r Subatomare Physik (SMI), Vienna, Austria\\
$^{102}$ SUBATECH, IMT Atlantique, Nantes Universit\'{e}, CNRS-IN2P3, Nantes, France\\
$^{103}$ Sungkyunkwan University, Suwon City, Republic of Korea\\
$^{104}$ Suranaree University of Technology, Nakhon Ratchasima, Thailand\\
$^{105}$ Technical University of Ko\v{s}ice, Ko\v{s}ice, Slovak Republic\\
$^{106}$ The Henryk Niewodniczanski Institute of Nuclear Physics, Polish Academy of Sciences, Cracow, Poland\\
$^{107}$ The University of Texas at Austin, Austin, Texas, United States\\
$^{108}$ Universidad Aut\'{o}noma de Sinaloa, Culiac\'{a}n, Mexico\\
$^{109}$ Universidade de S\~{a}o Paulo (USP), S\~{a}o Paulo, Brazil\\
$^{110}$ Universidade Estadual de Campinas (UNICAMP), Campinas, Brazil\\
$^{111}$ Universidade Federal do ABC, Santo Andre, Brazil\\
$^{112}$ Universitatea Nationala de Stiinta si Tehnologie Politehnica Bucuresti, Bucharest, Romania\\
$^{113}$ University of Derby, Derby, United Kingdom\\
$^{114}$ University of Houston, Houston, Texas, United States\\
$^{115}$ University of Jyv\"{a}skyl\"{a}, Jyv\"{a}skyl\"{a}, Finland\\
$^{116}$ University of Kansas, Lawrence, Kansas, United States\\
$^{117}$ University of Liverpool, Liverpool, United Kingdom\\
$^{118}$ University of Science and Technology of China, Hefei, China\\
$^{119}$ University of South-Eastern Norway, Kongsberg, Norway\\
$^{120}$ University of Tennessee, Knoxville, Tennessee, United States\\
$^{121}$ University of the Witwatersrand, Johannesburg, South Africa\\
$^{122}$ University of Tokyo, Tokyo, Japan\\
$^{123}$ University of Tsukuba, Tsukuba, Japan\\
$^{124}$ Universit\"{a}t M\"{u}nster, Institut f\"{u}r Kernphysik, M\"{u}nster, Germany\\
$^{125}$ Universit\'{e} Clermont Auvergne, CNRS/IN2P3, LPC, Clermont-Ferrand, France\\
$^{126}$ Universit\'{e} de Lyon, CNRS/IN2P3, Institut de Physique des 2 Infinis de Lyon, Lyon, France\\
$^{127}$ Universit\'{e} de Strasbourg, CNRS, IPHC UMR 7178, F-67000 Strasbourg, France, Strasbourg, France\\
$^{128}$ Universit\'{e} Paris-Saclay, Centre d'Etudes de Saclay (CEA), IRFU, D\'{e}partment de Physique Nucl\'{e}aire (DPhN), Saclay, France\\
$^{129}$ Universit\'{e}  Paris-Saclay, CNRS/IN2P3, IJCLab, Orsay, France\\
$^{130}$ Universit\`{a} degli Studi di Foggia, Foggia, Italy\\
$^{131}$ Universit\`{a} del Piemonte Orientale, Vercelli, Italy\\
$^{132}$ Universit\`{a} di Brescia, Brescia, Italy\\
$^{133}$ Variable Energy Cyclotron Centre, Homi Bhabha National Institute, Kolkata, India\\
$^{134}$ Warsaw University of Technology, Warsaw, Poland\\
$^{135}$ Wayne State University, Detroit, Michigan, United States\\
$^{136}$ Yale University, New Haven, Connecticut, United States\\
$^{137}$ Yildiz Technical University, Istanbul, Turkey\\
$^{138}$ Yonsei University, Seoul, Republic of Korea\\
$^{139}$ Affiliated with an institute formerly covered by a cooperation agreement with CERN\\
$^{140}$ Affiliated with an international laboratory covered by a cooperation agreement with CERN.\\

\end{flushleft} 
  %%%%%%% done by webmaster team                                                                            
\end{document}